# A Novel Magnetic Respiratory Sensor for Human Healthcare

Kee Young Hwang[1], Valery Ortiz Jimenez[1], Baleeswaraiah Muchharla[1], Tatiana Eggers[1], Anh-Tuan Le[2], Vu Dinh Lam[3], and Manh-Huong Phan[1,*]

[1]Department of Physics, University of South Florida, Tampa, Florida 33620, USA
[2]Phenikaa University Nano Institute (PHENA), Phenikaa University, Hanoi 12116, Vietnam
[3]Gradduate University of Science and Technology (GUST), Vietnam Academy of Science and Technology, 18 Hoang Quoc Viet, Ha Noi, Vietnam

Breathing is vital to life. Therefore, the real-time monitoring of a patient's breathing pattern is crucial to respiratory rehabilitation therapies such as magnetic resonance exams for respiratory-triggered imaging, chronic pulmonary disease treatment, and synchronized functional electrical stimulation. While numerous respiratory devices have been developed, they are often in direct contact with a patient, which can yield inaccurate or limited data. In this study, we developed a novel, non-invasive, and contactless magnetic sensing platform that can precisely monitor a patient's breathing, movement, or sleep patterns, thus providing efficient monitoring at a clinic or home. A magneto-LC resonance (MLCR) sensor converts the magnetic oscillations generated by a patient's breathing into an impedance spectrum, which allows for a deep analysis of one's breath variation to identify respiratory-related diseases like COVID-19. Owing to its ultrahigh sensitivity, the MLCR sensor yields a distinct breathing pattern for each patient tested. The sensor also provides an accurate measure of the strength of a patient's breath at multiple stages as well as anomalous variations in respiratory rate and amplitude. This suggests that the MLCR sensor can detect symptoms of COVID-19 in a patient due to shortness of breath or difficulty breathing as well as track the disease's progress in real time.

*Corresponding author: phanm@usf.edu (M.H.P)

## 1. Introduction

Bacteria and viruses threaten our lives on a daily basis and are responsible for severe plagues. To prevent further infection, it is essential to develop high-performance healthcare monitoring devices that can detect a pathogen and provide accurate health condition data and progress [1,2]. For instance, the development of novel respiratory devices to monitor breathing has advanced healthcare service, especially for those experiencing asthma or chronic obstructive pulmonary diseases.

In December 2019, the severe acute respiratory syndrome coronavirus 2 (SARS-CoV-2), also known as COVID-19, began to spread across the globe and has since caused the suffering and loss of life of millions of people [3,4]. This deadly virus has a structure similar to severe acute respiratory syndrome (SARS) and is highly contagious as it can spread from one person to another through air [5]. To limit the spread of COVID-19 and to help doctors in clinical decision making, it is critical to have fast screening and real-time monitoring of symptoms and growth at all stages (early, intermediate, and severe). Although detecting and tracking COVID-19 represents an incredibly challenging task, recent efforts have been devoted to achieving this ultimate goal [6-13].

One well-known method to detect the virus is based on the reverse transcriptase quantitative polymerase chain reaction (RT-qPCR) [6,7]. By collecting samples from subjects and using the reverse transcriptase enzyme, a complementary viral DNA (cDNA) is created from the RNA of the virus. Some portions of cDNA go through a polymerase chain reaction and can thus be detected after an amplification process [8]. Another method for detecting COVID-19 uses an electronic sensor. A field-effect transistor (FET), which can bind to COVID-19, has been developed and real-time changes to the electric signal have been demonstrated [9]. Optical biosensors and biochemical tests have also been employed to detect this deadly virus [10-12]. Despite the advantages they provide, these sensors are vulnerable to human contact, inaccuracy with small amounts of sample, and usually have long processing times [13].

Common symptoms of COVID-19 include: (i) shortness of breath or difficulty breathing, (ii) cough, and (iii) fever. Some patients suffer symptoms similar to lung disease such as pneumonia [5]. During the infection period, a patient likely experiences shortness of breath or difficulty breathing, which can be detected using a real-time respiratory monitoring device [14-16]. Current respiratory monitoring techniques are based on the impedance pneumography of the respiratory inductance plethysmography (RIP). By attaching electrodes to a patient's chest, the expansion and contraction of the chest causes a change in electrical impedance that can be expressed as the motion of respiration [15]. However, this contact-based method comes with limited sensitivity and reliability. It is also an inconvenient method to perform actual tests, such as children's respiratory monitoring. Therefore, there is an urgent need for the development of contactless devices that enable early and fast detection of COVID-19 or related diseases and track their growth rates in real time.

In this paper, we present a novel, non-invasive, and contactless magnetic sensing platform that integrates an ultrasensitive magneto-LC resonance (MLCR) sensor with a tiny permanent magnet for applications in healthcare monitoring. We demonstrate how to use this sensing platform to monitor respiratory motion (frequency, amplitude, shortness of breath, etc.) in real time and the analysis of patients' breathing pattern at different ages, at multiple stages, after holding breath, in both wake-up and sleep modes. Based on these findings, we propose an innovative approach that integrates magnetic sensing with machine learning for the fast screening and real-time tracking of COVID-19 and related viruses.

## 2. Magneto-LC Resonance Based Respiratory Monitoring

The MLCR sensor that utilizes the advantages of the giant magneto-impedance (GMI) effect of an ultrasoft magnetic Co-rich microwire [16] and an LC resonance circuit of an inductive coil made of the Co-rich microwire has recently been developed by our group for real time respiratory monitoring [17] and for spintronics applications [18]. Details of the design and fabrication of the MLCR sensor have been reported in Ref. [17]. In this study, we aimed to optimize the sensing capacity of the MLCR sensor. To do so, we performed a systematic investigation on the sensitivity of the MLCR sensor as a function of the operation frequency and the distance between the sensor and the targeted sensing object. To determine the optimized operation frequency and distance from the MMC sensor to a tiny permanent magnet, a vibrator with a non-magnetic arm was used. The waveform driving the vibrator was sinusoidal wave with 0.1 Hz frequency with 3.3 V amplitude, which was 0.5 cm from top to bottom. Changes in the external magnetic field, due to vibration of the permanent magnet, drives a change in the impedance of the MLCR sensor, corresponding to the highest and lowest peaks, which is defined as the sensitivity of the MLCR sensor (i.e. sensitivity = $Z_h - Z_l$, where $Z_h$ and $Z_l$ are the highest and lowest values of impedance). Figure 1a displays a typical vibrating (or breathing) pattern measured by the MLCR sensor as it detects the magnetic signal of a vibrating permanent magnet that was located 5 cm away from the sensor. By studying the effect of variation of operation frequency from 60 to 140 MHz on the sensor's sensitivity, we found the optimal operation frequency of 80 MHz to achieve the highest sensitivity (see Fig. 1b). Using this optimal operation frequency, we investigated the effect of changing the distance between the sensor and the permanent magnet on the sensor sensitivity and found that the 5 cm distance is optimal for magnetic respiratory monitoring using the MLCR sensor (see Fig. 1c). Therefore, in the thesis research, we operated the MLCR sensor at $f = 80$ MHz and placed it at $d = 5$ cm from the permanent magnet for all breathing monitoring experiments.

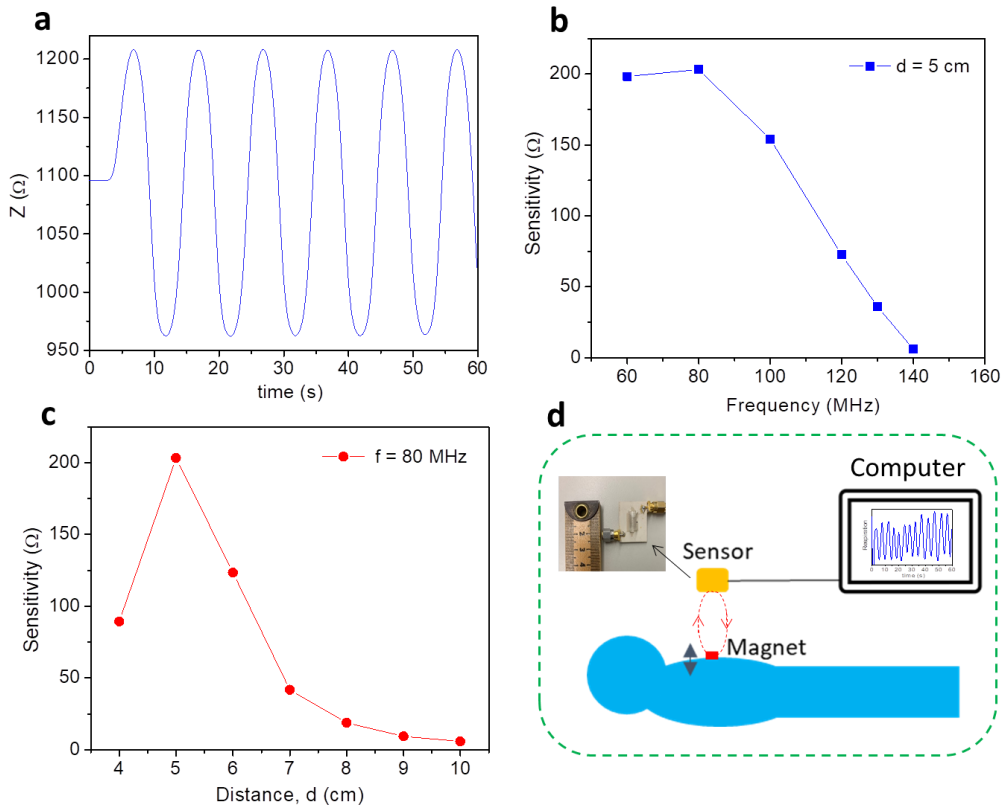

**Fig. 1** (a) Real-time monitoring of a vibrator's motion with a vibration frequency of 0.1 Hz and an optimal distance of 5 cm from a tiny permanent magnet. This simulates the normal breathing patterns of a patient; (b) The sensitivity of the MLCR sensor versus frequency at a distance 5 cm away from a vibrating tiny permanent magnet; (c) the sensitivity of the MLCR sensor versus distance (*d*) for $f = 80$ MHz; (d) a patient lays down on a comfortable bed for real-time respiratory monitoring.

To examine the sensing capacity and consistency of the MLCR sensor in monitoring human respiratory motion in real time, we designed experiments with the sensor held on a non-magnetic arm at a distance of 5 cm away from a tiny permanent magnet (e.g., NdFeB) mounted to the skin of a test subject. As the test subject breathes in and out, the chest will rise and fall, which generates a corresponding magnetic field change of the permanent magnet tracked by the MLCR. In other words, the motion of the permanent magnet is converted into an impedance change of the MLCR sensor. To perform human respiratory motion tests, we have designed a comfortable test configuration as illustrated in Fig. 1d.

### 3. Real-Time Human Healthcare Monitoring

*3.1. Real-time monitoring of human respiratory motion at different ages*

Research has shown that respiratory rate or breathing frequency decreases with age [19-25]. For instance, an infant (1 to 12 moths) breathes between 30 to 60 times per minute, toddlers (1-2 years) breathe anywhere from 24 to 40 times per minute, and preschoolers (3-5 years) breathe between 22 to 34 times per minute. While school-aged children (6-18 years) breathe anywhere from 18 to 30 times per minute, an adult (18-35 years) breathes between 12 to 18 times per minute. Elderly people (> 50 years) have respiratory rates between 16 and 25 breaths per minute. A respiratory rate of more than 25 breaths per minute may indicate a respiratory issue such as tachypnea [24,25].

Therefore, it is essential to examine if the magnetic respiratory device can precisely monitor and distinguish breathing patterns of people at different ages. The tests were performed on three volunteered subjects (a, b, c) of ages 30, 36, and 42 years old (YO). The results shown in Fig. 2(a-c) indicated that the 30 YO breathed the strongest and at a rate of 13 times per minute. The 36 YO breathed 15 times per minute and the 42 YO breathed 17 times per minute. The tests were repeated several times and showed reproducible and consistent results (see Fig.

2(d-f)). These results demonstrate that each person has a characteristic breathing pattern. When one gets older, they breathe weaker and the number of breaths per minute increases. This finding is consistent with those reported in previous works [19-25].

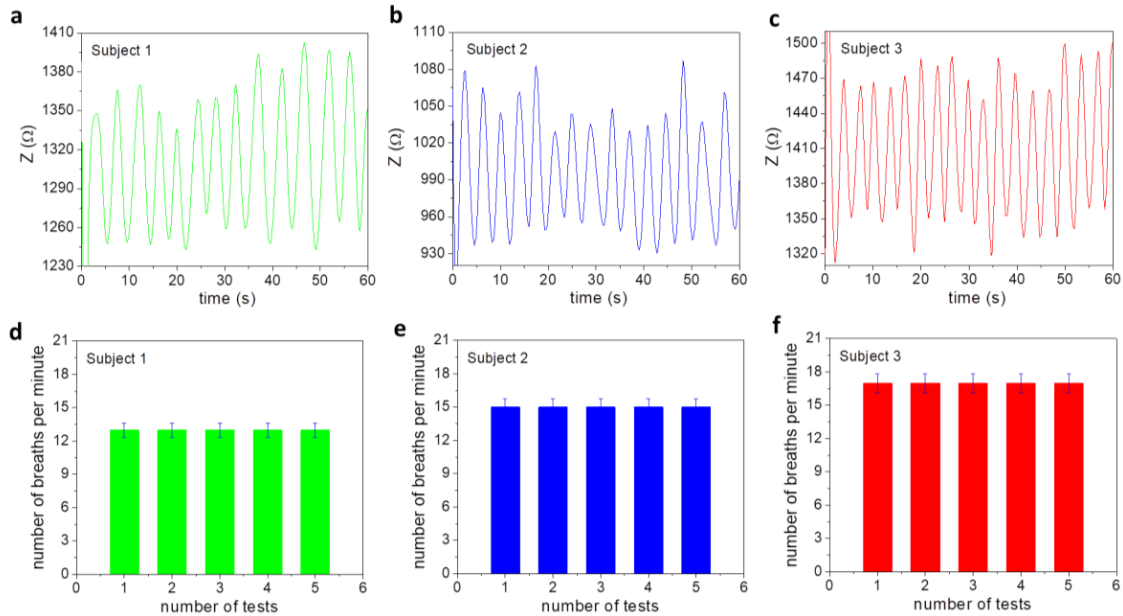

**Fig. 2** Normal breathing patterns of three subjects of ages (a) 30, (b) 36, and (c) 42 years old. Tests were repeated five times and showed consistent and reproducible results in number of breaths per minute for each subject (d-f).

Shortness of breath or abnormal breathing is noticed near $t = 20$ s for the 30 YO, between 20-40 s for the 36 YO, and at 35 and 45 s for the 42 YO in Fig. 2(a-c). To gain more information from these features, we analyzed the peak-to-peak time and plotted it as a function of time. As one can see in Fig. 3(a-c), there are some abnormal changes in the breathing pattern at the previously mentioned times. Such features of abnormal breathing or shortness of breath are not detectable by an Apple Watch tracker or a commercially available respiratory monitor. This demonstrates the increased capacity of our ultrasensitive magnetic respiratory device for human healthcare monitoring.

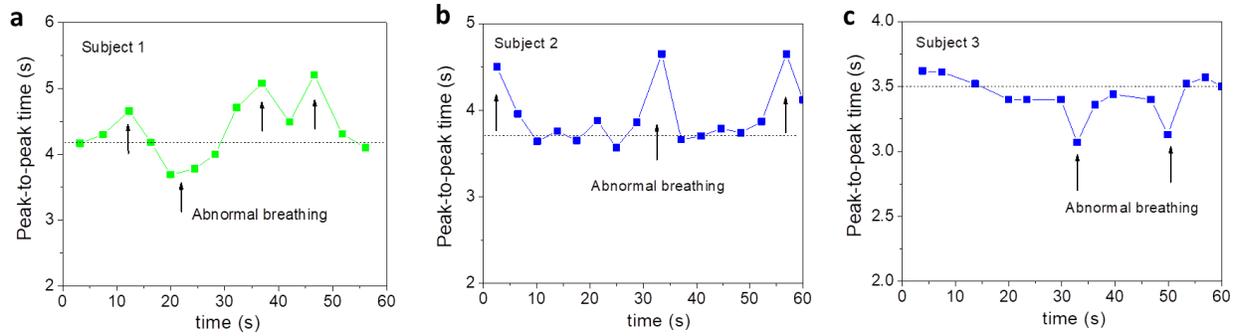

**Fig. 3** Peak-to-peak time as a function of time for three subjects aged (a) 30, (b) 36, and (c) 42 years old. These abnormal breathing patterns are related to those seen in **Fig. 2**.

*3.2 Real-time monitoring of human respiratory motion in multiple breathing states*

In practical applications, it is essential that a sensor is able to precisely monitor a patient's breathing in various and complex states in real time [24,25]. Existing respiratory monitoring devices are unable to provide this information. In this study, we tested the ability of the MLCR sensor to monitor a patient's respiratory motion in simple and complex breathing states. Subject 1 was instructed to breathe in two different states, namely, the normal and deep breathing states. The MLCR sensor precisely tracked how his/her breath (breathing amplitude and rate) varied and transitioned from one state to another (Fig. 4), which demonstrates the excellent capacity of the MLCR sensor for real-time tracking of a patient's health progress.

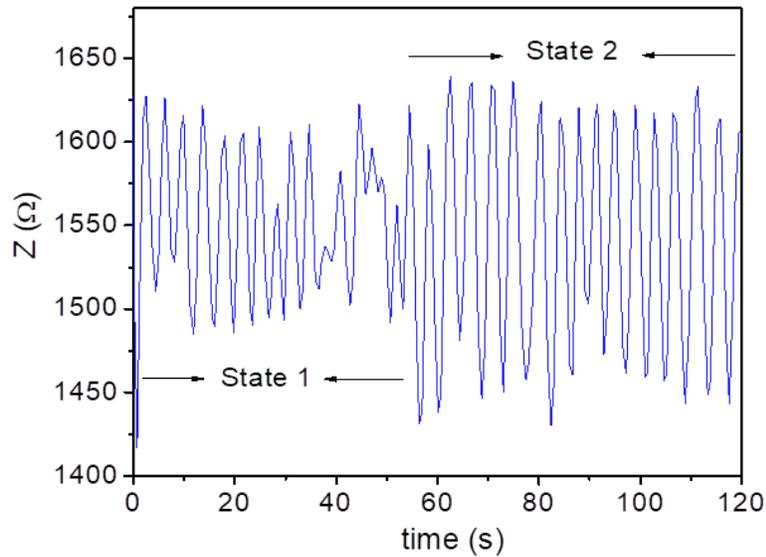

**Fig. 4** Breathing patterns of Subject 1 (30 YO) in normal (State 1) and deep (State 2) breathing states. The sensor provided an accurate measure of the strength of breath at these two states. An obvious breathing transition from the first to the second state is observed.

It has been reported that breathing training can significantly enhance health and physical performance in humans [24]. A proper breath holding technique can increase the amount of carbon dioxide in our cells and help us reveal personal issues we need to confront [25,26]. It has been shown that breath control improves lung function, which in turn preserves the health of stem cells and improves longevity, as well as promotes the regeneration of new tissues in the brain [27]. It has been reported that daily exercise of breath control can positively impact the sympathetic nervous system and the immune system by increasing the production of anti-inflammatory mediators [24-26].

In this regard, we explored the ultrahigh sensitivity of the MLCR sensor for real-time tracking of a patient's breathing in multiple/complex states from holding breath (State 1) to the release of breath (State 2) and the return to normal breathing (State 3). A clear understanding of this complex breathing process provides valuable information of the patient's health and may help doctors to develop an appropriate treatment plan for the patient. In our study, subjects 2 and 3 were instructed to first hold their breath as long as they could (State 1), then release their breath (State 2), and finally come back to normal breathing (State 3). The selected test results are displayed in Fig. 5.

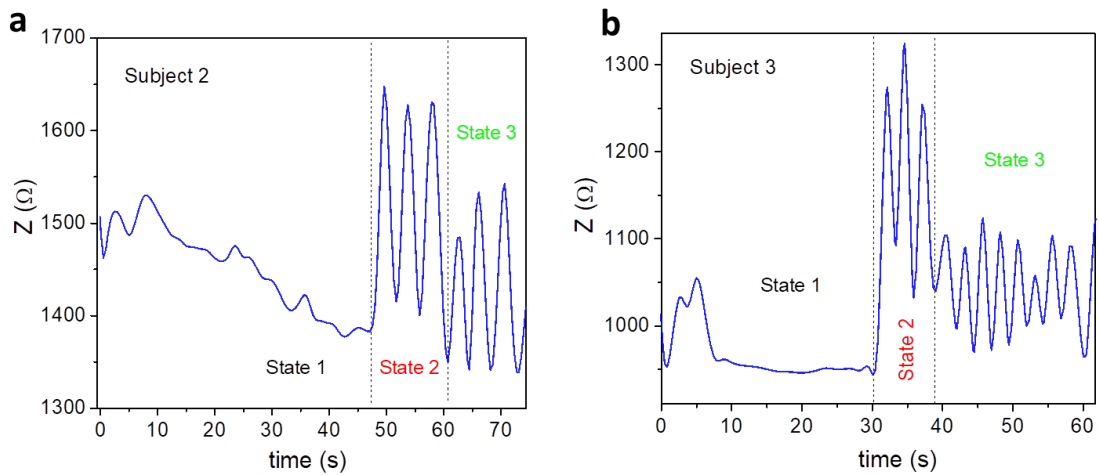

**Fig. 5** Breathing patterns of subjects aged (a) 36 and (b) 42 years old at three different states; State 1: breath holding, State 2: breath releasing, and State 3: normal breathing.

It is worth noticing in this figure that both the subjects underwent three distinguishable states, which were precisely tracked by the MLCR sensor. Subject 2 held the longest breath for 47 s (State 1) and, after releasing the breath, took 15 s to catch their breath (State 2) before returning to normal breathing (State 3). Meanwhile, Subject 3 held their longest breath for 30 s (State 1) and, after releasing the breath, took only 10 s to catch their breath (State 2) before returning to normal breathing (State 3). Compared to Subject 3, Subject 2 breathed more regularly after releasing their breath toward the normal breathing state. These findings seem to suggest that when getting older, one's ability to hold his/her breath decreases to a shorter period of time and they may experience more respiratory issues. Using this testing method, one can be aware of their current health status and be encouraged to follow an appropriate breathing control method to improve health and physical performance.

*3.3 Real-time monitoring of human respiratory motion in wake-up and sleep modes*

Nowadays, many people experience insomnia, sleep difficulties, dream sleep, or chronic trouble falling asleep. According to the American Sleep Association (ASA), insomnia is the most common sleep disorder [28]. Currently, about 30 percent of American adults are facing with insomnia. Insomnia can arise for various reasons such as irritability, depression or anxiety, stress from multiple work tasks, medical conditions, or mental health disorders and it can reduce our health, work performance, and quality of life. Research has shown that breathing training techniques can help us relax, sleep, and breathe more naturally and effectively [24,29]. Before using such techniques, it is essential to understand what affects our breathing before, during, and after our sleep [29]. In this context, there is an urgent need for the development of a highly sensitive, contactless respiratory monitoring device that can precisely monitor human respiratory motion in real time in these states. Owing to its ultrahigh sensitivity, the MLCR sensor is expected to fulfil this requirement.

In this study, we examined the capacity of the MLCR sensor in monitoring a patient's respiratory motion during sleep, as well as tracking a transition of his/her respiratory motion from wake-up mode to sleep mode. First, we show how respiratory motion of a patient can be tracked in real time during sleep by the MLCR sensor. In this case, the breathing pattern of Subject 2 (36 YO) was tracked while the subject was sleeping. As one can see clearly in Fig. 6, the sleeping patterns were clearly recorded. It is worth noticing that an abnormal breathing or a shortness of breath occurred around $t = 80$ s. This may suggest that this subject experiences some respiratory problems or associated diseases.

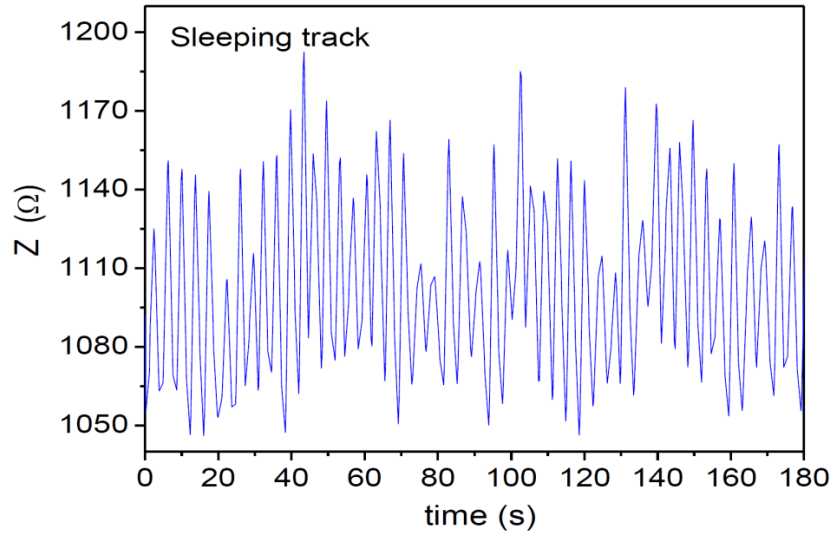

**Fig. 6** Breathing pattern of Subject 2 (36 YO) during 180 s of sleep and periodic sleeping patterns were observed.

Next, we tracked how the respiratory motion of a patient (Subject 1 and Subject 3) changed during their wake-up or sleep period, as well as from the transition from wake to sleep mode. The breathing patterns of Subject 1 (30 YO) were continuously tracked from wake to sleep mode, the result of which is shown in Fig. 7. This figure clearly shows a clear breathing transition of the subject from the wake-up to the sleep mode. It is also worth noting that in sleep mode, the subject breathed more deeply (higher amplitude) and slowly (11 times per minute) as compared to the wake mode (12 times per minute). In a similar fashion, the breathing patterns of Subject 3 (42 YO) were continuously tracked from wake to sleep mode (see Fig. 8). The sensor precisely tracked how his breath transitioned from wake to sleep mode. In sleep mode, the subject breathed more deeply (higher amplitude) and slowly (14 times per minute) as compared to the wake mode (20 times per minute). The results obtained from both Subject 1 and Subject 2 consistently pinpoint the fact that, in a normal sleep mode, one tends to relax and sleep better than during the wake mode. Relative to Subject 1, there is more of a difference in the respiratory rate between the wake and sleep mode found for Subject 3. This suggests that older people tend to relax more during sleep [29].

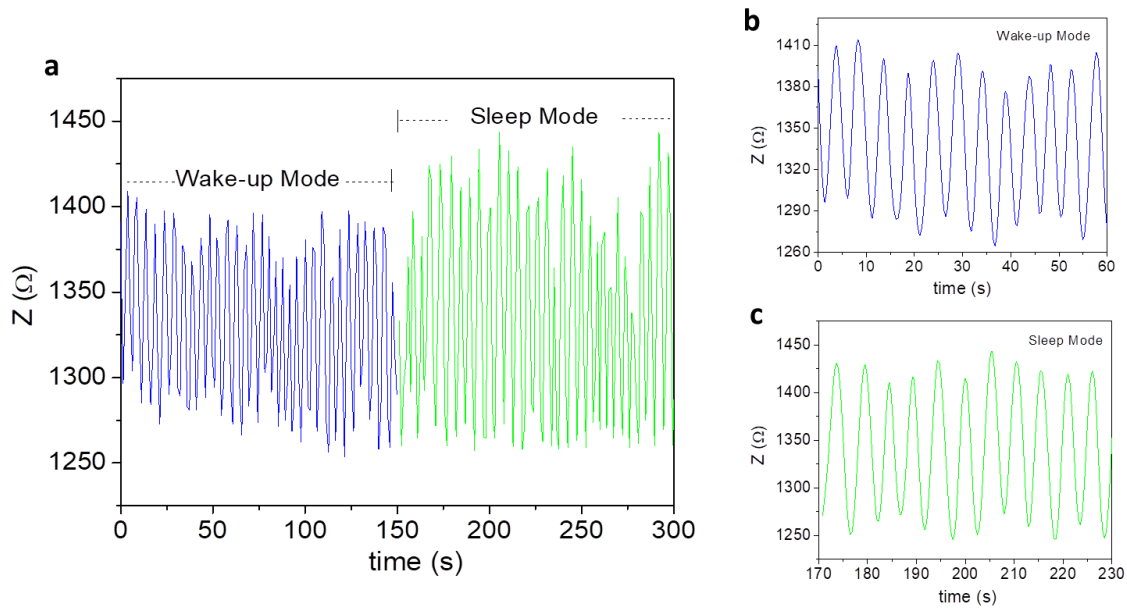

**Fig. 7** (a) Breathing patterns of Subject 1 in both wake and sleep modes. An obvious transition from the wake to sleep mode was observed. (b) and (c) show the wake and sleep breathing patterns over 1 min.

In addition, the ultrahigh sensitivity of the MLCR sensor enables the detection of abnormal breathing around $t = 40$ s in both the wake-up and sleep breathing patterns for Subject 3. This indicates that the noticed abnormal breathing is intrinsic to Subject 2 and that this subject should be aware of this potential respiratory issue or related disease (e.g., chronic pulmonary disease) and should thus perform daily breathing training/exercises to resolve it [24,29].

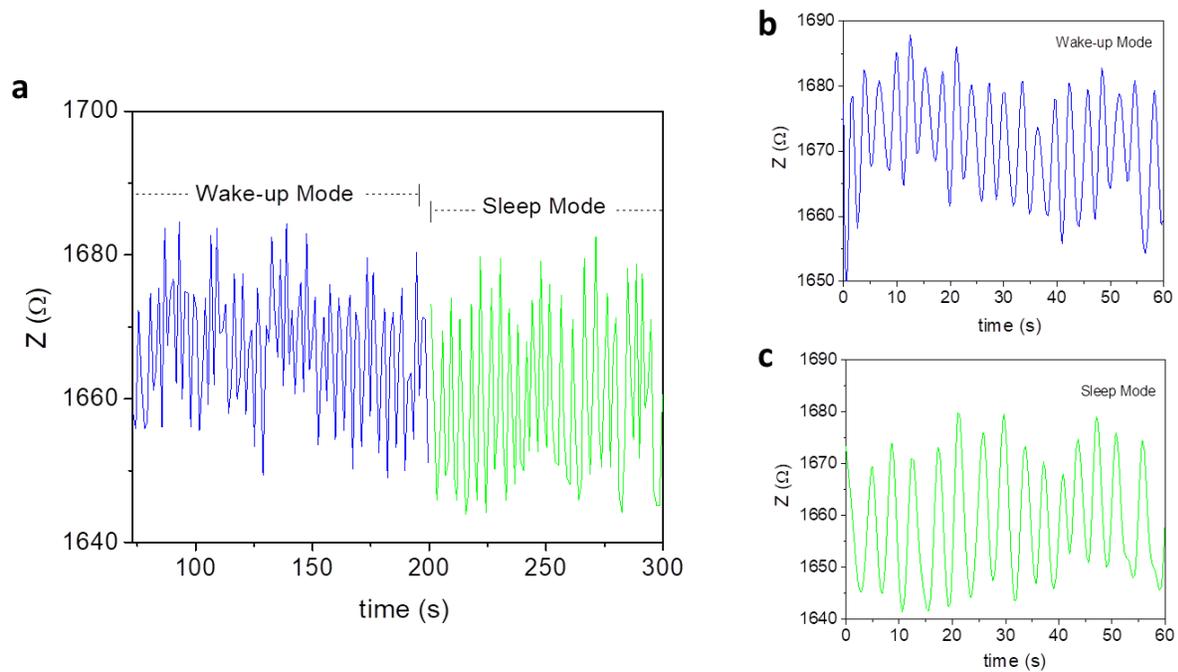

**Fig. 8** (a) Breathing patterns of Subject 3 in both wake and sleep modes. An obvious transition from the wake to the sleep mode was observed. (b) and (c) show the wake-up and sleep breathing patterns per 1 min.

## 4. Summary and Future Outlook

We have developed an innovative, contactless magnetic respiratory monitoring platform based on ultrasensitive magneto-LC resonance (MLCR) technology and a permanent magnet. This technology has been successfully applied to test human respiratory motion in real time and in different states. We demonstrated that the MLCR sensor can track a person's breathing pattern very precisely in real time over the course of multiple states (normal breathing, holding breath, and sleeping). Owing to its ultrahigh sensitivity, the sensor yielded a distinct breathing pattern for each person tested and revealed abnormal breathing, which is superior to existing respiratory monitoring devices. We have shown that when getting older, one usually manifests weaker breathing marked by an increasing breathing rate. He/she tends to hold his/her breath for a shorter period of time and often experiences more with respiratory issues than younger people. People become more relaxed, thus breath more regularly and slowly while sleeping as compared to the wake-up mode. Our highly sensitive magnetic respiratory monitor provides not only valuable information on a patient's current health status, but also a novel breathing control tool for improving our health and physical performance.

Through these studies, we also identify several exciting and important applications of our novel magnetic respiratory monitoring technology for advanced human healthcare systems. In particular, research on the practical implementation of this technique for the fast screening and real-time tracking of COVID-19 or associated viruses is very exciting and worth pursuing. Monitoring COVID-19 patients remotely with the MLCR detector and collecting data from numerous physiological signals, with the help of machine learning, would improve clinical decision making for providers [29-32]. The Fast Fourier Transform (FFT) algorithm, which converts a signal from the time domain to the frequency domain and vice versa, can be used to extract spectral features from breathing patterns. This can also be used by machine learning and other algorithms for signal processing and to provide comprehensive information on a patient's health status and physical performance.

## Acknowledgements

Research at USF was supported by the USF Covid19 Research Foundation under Grant number 125300. The work was also supported by Vietnam Ministry of Science and Technology through the national-level project ĐTĐLCN.17/19.